\documentclass[12pt,fleqn]{article}
\usepackage{amsmath,a4,latexsym,epsfig}
\newcommand{\intd}{\int \! d^4 x \;}

\newcommand{\Ga} {\Gamma}
\newcommand{\Gacl} {{\Gamma_{\rm cl}}}

\newcommand{\lambdabar}{{\overline\lambda}}
\newcommand{\sigmabar}{{\overline\sigma}}
\newcommand{\psibar}{{\overline\psi}}

\newcommand{\phibar}{{\overline \varphi}}
\newcommand{\epsilonbar}{{\overline\epsilon}}
\newcommand{\thetabar}{{\overline\theta}}
\newcommand{\etabar}{{\overline\eta}}
\newcommand{\chibar}{{\overline\chi}}
\newcommand{\qbar}{{\overline q}}
\newcommand{\fbar}{{\overline f}}
\newcommand{\cbar}{{\overline c}}

\newcommand{\alphadot}{{\dot\alpha}}

\newcommand{\lambdaV}{{\tilde \lambda}}
\newcommand{\lambdaVbar}{{\overline {\tilde \lambda}}}
\newcommand{\DV}{{\tilde D}}

\def\dF#1{\frac{\delta{\cal F}}{\delta#1}}
\def\dfunc#1#2{\frac{\delta^{#1}}{\delta#2}}

\def\pslash#1{{\setbox0=\hbox{$#1$}
  \rlap{\ifdim\wd0>.7em\kern.22\wd0\else\kern.1\wd0\fi /}#1}}

\def\brs{\mathbf s}
\newcommand{\mn}{\mu\nu}

\begin{document}
\begin{titlepage}

\begin{flushright}
KA--TP--22--2001\\
BN--TH--01--05\\
{\tt hep-ph/0107061}\\
\end{flushright}

\begin{center}
{\Large\bf{
        Non-renormalization theorems  in   softly broken\\[1ex]
         SQED and the soft  $\beta$-functions}}
\\
\vspace{8ex}
{\large       E. Kraus$^a$} and
{\large        D. St{\"o}ckinger$^b$}
{\renewcommand{\thefootnote}{\fnsymbol{footnote}}
\footnote{E-mail addresses:\\
                kraus@th.physik.uni-bonn.de,\\
                ds@particle.physik.uni-karlsruhe.de.}} 
\\
\vspace{2ex}
{\small\em               $^a$ Physikalisches Institut,
              Universit{\"a}t Bonn,\\
              Nu{\ss}allee 12, D--53115 Bonn, Germany\\
}
\vspace{.5ex}
{\small\em $^b$ Institut f{\"u}r Theoretische Physik, 
              Universit{\"a}t Karlsruhe,\\
              D--76128 Karlsruhe, Germany\\}
\vspace{2ex}
\end{center}
\vfill
{\small
 {\bf Abstract}
 \newline\newline
The renormalization of softly broken SQED is related to the one of
supersymmetric
QED by using  the construction   with a local
gauge supercoupling and by taking into account softly broken anomalous
axial $U(1)$ symmetry. From this extended model one obtains the
non-renormalization theorems of SQED and 
 the counterterms of the soft breaking
parameters as functions of the supersymmetric counterterms.
 Due to
the Adler--Bardeen anomaly of the axial current an invariant
regularization scheme
does not exist, and therefore
the  $\beta$-functions of soft  breaking parameters   are
derived from an algebraic construction  of the Callan--Symanzik
equation and of the renormalization group equation. 
 We obtain the soft
$\beta$-functions in terms  of the gauge $\beta$-function and of the
anomalous dimension of the supersymmetric matter mass.
 In particular, we find that the $X$-term
of  the scalar mass $\beta$-function as well as the gauge
$\beta$-function in $l\geq 2$
are  due to  the Adler--Bardeen anomaly of the axial symmetry.

\vfill

}

\end{titlepage}


\newpage
\section{Introduction}

Starting with the investigations of Yamada \cite{YA94} there has been
considerable pro\-gress in relating the $\beta$-functions of soft
supersymmetry breaking parameters \cite{GIGR82} to the $\beta$-functions and
anomalous dimensions of the supersymmetric model
\cite{AKK98,KAVE00,JAJO97gaugino,JAJO98scalar2}. 

There are two different approaches in extracting $\beta$-functions of
the softly broken model from the respective supersymmetric model: The first
approach \cite{YA94, AKK98,  KAVE00} is based on the superfield formalism
 \cite{FULA75,GSR79,HE84,SCH85} and exploits
supergraph techniques to relate the UV counterterms of softly broken
and supersymmetric gauge theories. In the
second approach \cite{HISH98, JAJO97gaugino, JAJO98scalar, JAJO98scalar2}
 renormalization group invariant quantities are used for deriving
all-order expressions for the $\beta$-functions of soft breaking
parameters. The latter computation is mainly based on the component formalism
using dimensional reduction (DRED) as a supersymmetric regularization
scheme.

From the beginning it has been seen that the calculation of the scalar 
mass $\beta$-function is somewhat involved: In
dimensional reduction a mass term for $\epsilon$ scalars is
introduced, which enters  the $\beta$-function of the
scalar mass in two-loop order
\cite{JAJO94}.  By a redefinition from the DRED to a DRED$'$ scheme
the scalar mass
$\beta$-function can be defined in terms of physical parameters
\cite{VAMA94,JJMVY94}, and one
gets then an extra contribution, called $X$-term \cite{JAJO98scalar2}.
It is the $X$-term, in which the results 
 gained from a simple application of supergraph techniques
differ from the explicit expressions from 2-loop order onwards.

It is obvious that such ambiguities can be only resolved by
a scheme-independent construction of softly broken supersymmetric
models. In the present paper we introduce such a scheme-independent 
definition of softly broken SQED, which describes at the same time the 
 non-renormalization theorems and the
relation of soft breaking to the supersymmetric parameters. 
In particular it is
  shown that the $X$-term is  in effect
generated   by the supersymmetric extension of the
Adler--Bardeen anomaly. 


For the construction we use an extended model of SQED 
which  has been considered in a
recent paper \cite{KRST01} for the derivation of non-renormalization
theorems in the Wess--Zumino gauge. In addition to the
physical fields it contains  a local gauge coupling with its
superpartners, a chiral 
multiplet coupled to the mass term of matter and an axial vector
multiplet. In the extended model
 the photino mass is generated by the gauge supercoupling 
and the left-right mixing scalar mass is generated by
 the chiral field of the
matter mass term as proposed in the literature \cite{AKK98,
HISH98}. Differently from their approaches, the
scalar mass of matter  is generated by the $D$-component of the axial vector
multiplet.  The interaction of the axial vector multiplet with the
matter fields is governed by  softly broken axial symmetry. Axial
symmetry, however,
is broken by the Adler-Bardeen anomaly. In presence of the local gauge
coupling the Slavnov--Taylor
identity can be modified in such a way 
that it includes the anomaly. Then the renormalization of softly
broken SQED is defined by the anomalous Slavnov--Taylor identity.
From this identity we obtain the non-renormalization theorems in the
same way as
in the supersymmetric model, and 
 the relations between softly broken SQED and supersymmetric QED are derived
 from a scheme independent algebraic
construction. 


In this respect the  present construction is superior to the
constructions  \cite{MPW96b,HKS00} of
 softly broken supersymmetric theories in the
framework of algebraic renormalization.
 Irrespective of the fact whether the soft breakings are
introduced by BRS doublets  or by spurion fields, their  divergences
do not appear  to be related to the divergences of the supersymmetric
theories.   Like  the non-renormalization theorems,
the specific renormalization properties of softly broken supersymmetry
are not  direct consequence of  supersymmetry, but are a consequence of the
multiplet structure of supersymmetric Lagrangians \cite{FLKR00,
KRST01}.
 Indeed, the soft mass breakings are the lowest components of the
 Lagrangian multiplets and this property
is only exploited with the use of a local
supercoupling and of axial symmetry as proposed in the
present paper.

The plan of the paper is as follows: In section 2 we introduce the
classical action of the extended model of SQED including the soft
supersymmetry breaking.  In section 3 we give the defining symmetries
for the construction of higher-order Green functions. In particular, we
show that the Green functions satisfy an anomalous Slavnov-Taylor
identity, which includes the Adler Bardeen anomaly, to all orders.
In section 4 we derive the invariant counterterms and comment on the
non-renormalization theorems. In section 5 we discuss the
normalization conditions. Section 6 and section 7 is devoted to the
derivation of soft mass $\beta$-functions: We first derive the
Callan-Symanzik equation and prove then, that in mass independent
schemes their coefficient functions are related to the ones of the
renormalization group equation. The all-order expressions of soft
mass $\beta$-functions 
follow from symmetries
of the extended model in the same way as the closed form of the gauge 
$\beta$ function.
In the appendices we summarize the BRS transformations and the
symmetric operators in their general form. 
The   conventions are the ones we have given in appendix A of \cite{KRST01}.

\newpage

\section{The classical action}

In a recent paper \cite{KRST01}
the non-renormalization theorems of SQED have been
derived      in the Wess-Zumino gauge \cite{WZ74_SQED}
by extending the gauge
coupling to an external real superfield and by taking into account
softly broken axial symmetry with its anomaly.
The extended model of SQED  can be immediately used also for a consistent
description of the soft supersymmetry breaking.
The soft breaking terms   of the Giradello--Grisaru 
class \cite{GIGR82} are implicitly included in
the extended model since they couple to the highest components of external
field multiplets. 
They are generated when we introduce  a constant shift in these external
field components. Explicitely we introduce the following external field
multiplets with associated shifts: 
\begin{itemize}
\item The local gauge coupling $e(x)$ is the lowest component of a real
superfield
 $E (x, \theta, \thetabar )$, which itself is composed of a chiral and
 an
antichiral
field multiplet $\mbox{\boldmath{$\eta$}}(x, \theta)$ and
$\mbox{\boldmath{$\etabar$}} (x, \thetabar)$:
\begin{eqnarray}
\label{E2def}
E (x, \theta, \thetabar ) & = & ({\mbox{\boldmath{$\eta$}}}(x, \theta,
\thetabar ) +{\mbox{\boldmath{$\etabar$}}}(x, \theta, \thetabar ))^ 
{-\frac 12} \equiv e(x) + {\cal O}(\theta, \thetabar)
\end{eqnarray} 
with 
\begin{equation}
\label{defeta}
{\mbox{\boldmath{$\eta$}}}(x, \theta) = \eta + \theta^ \alpha
 \chi_\alpha + \theta^2 f \ ,
 \qquad 
{{\mbox{\boldmath{$\etabar$}}} }(x, \thetabar) = \etabar + \theta_\alphadot \chibar ^
 \alphadot  + \thetabar^2 {\overline f} 
\end{equation}
in the chiral and antichiral representation, respectively.
Shifting the highest components of the chiral and antichiral field multiplet
\begin{equation}
\label{fMlambda}
f \to  f + \frac {M_\lambda}{ e^ 2}\ , \qquad
\bar f \to  \bar f + \frac {M_\lambda}{ e^ 2} \ ,
\end{equation}
 we will obtain a mass term for the photino in the classical action.
\item   The axial current is coupled to an external axial vector multiplet.
When we  shift the $D$-component  of the axial vector multiplet
\begin{equation}
\label{Vdef}
V^ i =  (V^ {\mu},
 \lambdaV^ \alpha, \lambdaVbar^ \alphadot,\DV)
\end{equation}
with
\begin{equation}
\label{DM}
 \tilde D  \to   \tilde D - 2 M^2  \ ,
\end{equation}
a mass term for the scalar superpartners of the electron is present in
the classical action.
\item
 The chiral and antichiral field multiplets
\begin{equation}
\label{qdef}
q^ i =
(q,q^ \alpha, q_F)\quad \mbox{and} \quad
 \qbar^ i = (\qbar,\qbar^ \alpha, \qbar_F)
\end{equation} 
with dimension one
couple to the supersymmetric mass term.
By a shift in their $F$-components  
 one generates the left-right-mixing mass term, usually called
 the $b$-parameter,
 for the scalar superpartners
of the electron in the classical action:
\begin{equation}
\label{qFb}
q_F \to q_F - b \ , \qquad \qbar_F \to \qbar_F - b\ .
\end{equation}
The $b$-parameter has mass dimension 2.
\end{itemize}

The shifts appear not only in the classical action, but modify
 also the  axial transformations and  the supersymmetry
transformations of the corresponding fields. However, their presence
 in the symmetry transformations does not change
the algebraic characterization and the algebraic structure of symmetry
transformations and
 the model with soft breaking is
described by the same symmetries  as  the supersymmetric  model.
 Hence,   the 
 renormalization properties remain unchanged in  softly broken SQED.
 It is the purpose of the present paper to work out explicitly
the symmetric counterterms and the $\beta$-functions in presence of
soft supersymmetry breaking. All expressions are immediately obtained
 from the expressions in 
the symmetric model \cite{KRST01}, but for a clear presentation we summarize
the construction in a condensed form.

 In 
the Wess--Zumino gauge the algebra of supersymmetry transformations
closes on field dependent gauge
transformations and the gauge fixing of the photon cannot be given in
a supersymmetric form. To overcome these difficulties one uses a BRS
formalism and combines the symmetries of the model   in the
BRS-transformations \cite{White92a, MPW96a}. 
On fields with ghost charge zero the BRS operator
acts as a combination of gauge symmetry, axial symmetry, supersymmetry
and translations:
\begin{equation}
\label{BRS}
\brs \phi = (\delta^ {\rm gauge}_{c(x)} + \delta^ {\rm axial}_{\tilde
c(x)} + \epsilon ^ \alpha \delta_\alpha + \bar \delta_\alphadot
\epsilonbar^ \alphadot - i \omega^ \mu \partial_\mu) \phi \ .
\end{equation}
The ghost fields $c(x), \tilde c(x)$  replace the local
transformation parameters of gauge transformations and axial
transformations, and the constant ghosts $\epsilon^ \alpha $,
$\epsilonbar^ \alphadot$ and $\omega^ \mu$ are the constant supersymmetry
 and translational ghosts, respectively.
BRS-transformations of the ghosts are determined by the structure
constants of the algebra and the algebra of symmetry transformations
is expressed in on-shell nilpotency of the BRS operator. 
 The BRS transformations of the
fields are summarized in  appendix A. They differ from the BRS
transformations in the symmetric model by  shifts
in the scalar components of external fields.

The complete classical action 
 is  decomposed into the
physical  part $\Ga_{\rm {susy}}$,
 the 
  gauge fixing and ghost
part $\Ga_{\rm g.f}$ and  an external field part $\Ga_{\rm
ext.f}$, which makes possible to describe the BRS invariance of the action by
the Slavnov--Taylor identity:
\begin{equation}
\label{Gacl}
\Gacl = \Ga_{\rm {susy}} + \Ga_{\rm g.f.} + \Ga_{\rm ext. f.} \ .
\end{equation}
$\Ga_{\rm {susy}} $ is invariant under gauge transformations with
the local gauge coupling, supersymmetry
transformations and axial transformations. Up to normalization
constants it is determined by these symmetries and
the construction results in the following form:
\begin{eqnarray}
\label{Gasusy}
\Ga_{\rm susy} & = &  
\intd \Bigl( - \frac 1{4e^2}F^ {\mn}(eA) F_{\mn} (eA) + \frac i2 (
\lambda \sigma \partial \lambdabar - \partial \lambda \sigma
\lambdabar) \\
 & &\phantom{\intd}   +  \frac i 2 (\eta - \etabar ) \partial_\mu 
(e ^2 \lambda \sigma^ \mu \lambdabar)
- \frac i 8 (\eta - \etabar) \epsilon^{\mu \nu \rho\sigma} F_{\mn}(eA)
F_{\rho\sigma }(eA) \nonumber \\
& & \phantom{\intd}+ \frac i 4 e ( \chi \sigma^ {\mn} \lambda - \lambdabar {\bar
\sigma}^ {\mn} \chibar) F_{\mn}(eA)
-  \frac 12 e ^2 f \lambda \lambda - \frac 12  e ^2 {\overline f}
\lambdabar \lambdabar  \nonumber \\
& & \phantom{\intd} - \frac 12 M_\lambda (\lambda \lambda + \lambdabar
\lambdabar)
\nonumber \\
& & \phantom{\int} + \Bigl( 
D^ \mu  \phibar_L D_\mu \varphi_L+ i   \psi_L^ \alpha \sigma^ \mu
_{\alpha \alphadot}  D_\mu \psibar_L
^ {\alphadot} + i e Q_L  \sqrt{2}( \lambda \psi_L  \phibar_L  -
\lambdabar  \psibar_L 
 \varphi_L)   \nonumber
 \\
& & \phantom{\intd} 
+ i  \sqrt{2}( \lambdaV \psi_L  \phibar_L  -  
\lambdaVbar  \psibar_L
 \varphi_L) + \frac 1 2 \DV
\phibar_L \varphi_L + (_{L \to R}) \Bigr)\nonumber \\ 
& & \phantom{\intd} - M^2 (\phibar_L \varphi_L + \phibar_R \varphi_R)
  \nonumber \\
& & \phantom{\int}- \frac 18 \bigl( ie^2 (\chi \lambda - \chibar \lambdabar) + 2  e  Q_L (\varphi_L
\phibar _L -\varphi_R
\phibar _R )\bigr)^ 2 \nonumber \\
& & \phantom{\int}- (q +m )(\qbar+m) (\varphi_R \phibar_R  +  \varphi_L \phibar_L) -
\bigl((q+m) \psi_L^ \alpha \psi_{R \alpha} + \mbox{\rm c.c.}\bigr)\nonumber  \\
& & \phantom{\int} - \frac 1 {\sqrt 2}\bigl( q^ \alpha (\psi_{L\alpha} \varphi_R + 
\psi_{R\alpha} \varphi_L) +  \qbar_ \alphadot (\psibar_{L}^{\alphadot}
\phibar _R + 
\psibar_{R}^{\alphadot} \phibar_L)\bigr) \nonumber \\
& & \phantom{\int} + q_F \varphi_L \varphi_R + \bar q _F \phibar_L
\phibar_R  -b (\varphi_L \varphi_R +\phibar_L
\phibar_R \bigr) \Bigr) \nonumber \ .
\end{eqnarray}
The fields  $A^ \mu$ and $\lambda^ \alpha, \lambdabar^ \alphadot$
 are the photon and photino
 and $\varphi_A , \psi_A, A = L,R$ and their complex conjugate
 the left- and right-handed matter fields with
charge $Q_L = -1$ and $ Q_R = 1$.
The covariant derivatives are 
 covariant with respect to gauge transformations and  axial
 transformations:
\begin{equation}
\label{dcovariant}
D_\mu \phi_A = (\partial_\mu + i Q_A A_\mu + i V_\mu) \phi_A\ , \qquad
 D_\mu {\bar \phi}_A = ({D_\mu \phi_A})^ \dagger\ ,\quad
\phi = \psi, \varphi  \  .
\end{equation}

The gauge fixing and ghost part of the action, $\Ga_{\rm g.f.} $, as
 well as the external field part $\Ga_{\rm ext.f} $ 
 only depend on the local
gauge coupling $e(x)$ and have the same form
as in the symmetric model \cite{KRST01}.
 Using the auxiliary field  $B$ for describing the gauge
 fixing term, the gauge fixing and ghost part can be written as a
 BRS-variation:
\begin{equation}
\label{Gagf}
\Ga_{\rm g.f.} = \brs \intd \bigl( \frac 12
\xi \bar c B + \bar c \frac 1 e \partial(e A ) \bigr)
=\intd \bigl( \frac 12
\xi   B^ 2 + B \frac 1 e \partial(e A ) \bigr) + \Ga_{\rm ghost} \ .
\end{equation}
The  explicit form of $\Ga_{\rm ghost}$
 is not relevant for the further construction.


The  classical action satisfies the Slavnov-Taylor identity
\begin{equation}\label{ST}
{\cal S} (\Ga_{\rm cl}) = 0\ ,
\end{equation}
with the usual Slavnov-Taylor operator (see (\ref{STOperator})).
It describes   BRS invariance of the classical action as well as
 the
algebraic structure of symmetry transformations.

When we take the limit to constant coupling and set all further
external fields to zero, the classical action becomes the classical action of 
SQED with soft supersymmetry breaking:
\begin{equation}
\label{Gaconst}
\lim_{E \to e} \Ga_{\rm cl} \big|_{V^ i= 0 \atop q^ i =0} =
\Ga^ {\rm SQED}_{\rm cl} + \Ga_{\rm soft} \equiv \Ga_{\rm cl}^ {\rm sSQED}
\end{equation}
with
\begin{eqnarray}
\label{Gasoft}
\Ga_{\rm soft} &= &\intd \Bigl(-\frac 12 M_\lambda (\lambda \lambda +
\lambdabar \lambdabar) \nonumber \\
& & \phantom{\intd} - M^ 2 (\varphi_L \phibar_L + \varphi_R \phibar_R)
- b( \varphi_L \varphi_R + \phibar_L \phibar_R)\Bigr) \ .
\end{eqnarray}

The  classical action (\ref{Gacl}) is the starting point for the
perturbative calculations. It turns out that it is complete in the
sense of multiplicative renormalization.

 \section{Symmetries and renormalization}

 In the perturbative construction  the local
 coupling and its superpartners are considered as external fields which
 appear in the same way as ordinary external fields in the generating functional
 of 1PI Green functions $\Ga$. 
 However the chiral and antichiral fields which compose the local
 coupling
  are distinguished from  the dimensionless spurion fields
 by the property that the local gauge coupling is the perturbative
 expansion parameter. This is the content of 
 the topological formula 
 \begin{equation}
 \label{topfor}
 N_{e(x)} = N_{\rm amp. legs} + N_Y+ 2N_f+
 2N_\chi + 2 N_{\eta - \etabar} 
 + 2(l-1)\ ,
 \end{equation}
 which determines the number of local couplings in a specific diagram
 in dependence of the loop order $l$.
 Here  $N_{\rm amp. legs} $ counts the number of external
 amputated legs with propagating fields ($A^ \mu, \lambda, \varphi_A,
 \psi_A, c, \cbar  $ and the respective
 complex conjugate fields), $N_Y$ gives the number of BRS insertions,
 counted by the number  of differentiations with respect the the
 external fields $Y_\phi$. 
  $N_f$, $N_\chi$ and $N_{\eta - \etabar } $ gives the number of 
 insertions 
 corresponding to  the respective external fields. 
 As for the classical action the validity of the topological formula
 ensures that the limit to constant coupling results in the 1PI Green
 functions of ordinary SQED with soft breaking.

 The topological formula is not the only restriction on the appearance
 of the ${{\mbox{\boldmath{$\eta$}}} }$ and
 ${{\mbox{\boldmath{$\etabar$}}} }$-multiplets, but it is seen that the
 classical action depends on the parity odd scalar field $\eta -
 \etabar$ only via a total derivative. The corresponding Ward identity
 \begin{equation}
 \label{holomorph}
 \intd \bigl(\frac {\delta} {\delta\eta} - \frac {\delta}
 {\delta\etabar}\bigr)
 \Ga = 0 
 \end{equation}
 can be maintained in the course of renormalization. It was shown that
 this identity in combination with supersymmetry is the basis for the
 non-renormalization theorems \cite{KRST01}.

 Supersymmetry, abelian gauge symmetry and axial symmetry 
  are  included in the Slavnov--Taylor identity. However,
 at the quantum level  axial symmetry is broken by the
 Adler--Bardeen anomaly \cite{AD69,ADBA69}. In the  model with local gauge
 coupling the Adler-Bardeen anomaly can be absorbed into a modified
 Slavnov-Taylor identity and due to the non-renormalization of the
 anomaly  it
  can be proven that the generating
 functional of 1PI Green functions satisfies the following anomalous
 Slavnov--Taylor identity to all orders \cite{KRST01}:
 \begin{equation}
 \label{STanomalous}
  {\cal  S} (\Ga) + r^{(1)} \delta
 {\cal S} \Ga =  0 \ ,
 \end{equation}
 where ${\cal S}(\Ga)$ is defined in (\ref{STOperator}) and
 \begin{eqnarray}
 \label{STanomalousoperator}
 \delta {\cal S}\Ga 
 & = & - 4 i  \intd  \Bigl(\tilde c \Bigl(\frac {\delta}{\delta \eta} -
 \frac {\delta}{\delta \etabar}\Bigr) + 2i (\epsilon \sigma^ \mu )^
 \alphadot V_\mu \frac {\delta}{\delta \chibar^ \alphadot} 
  - 2i ( \sigma^ \mu \epsilonbar )^ \alpha V_\mu \frac
  {\delta}{\delta \chi^ \alpha} \nonumber \\  & & \phantom{\intd}
   + 2 \epsilonbar_\alphadot \lambdaVbar ^\alphadot 
 \frac {\delta}{\delta f} - 2 \lambdaV^\alpha \epsilon_\alpha \frac {\delta}{\delta
  \fbar} \Bigr) \Ga \ .
 \end{eqnarray}
 Here $r^ {(1)}$ is the coefficient of the anomaly determined from the
 usual triangle diagrams:
 \begin{equation}
 \label{r1}
 r ^{(1)} = - \frac 1 {16 \pi^ 2} \ .
 \end{equation}
 The operator (\ref{STanomalousoperator}) describes the supersymmetric
 extension of the Adler--Bardeen anomaly. Due to the appearance of the
 supersymmetry ghosts $\epsilon ^ \alpha$ and $\epsilonbar^ \alphadot$
 it modifies the supersymmetry transformations of the axial vector
 multiplet. It has been shown in ref.~\cite{KRST01} that these
 modifications are the Wess-Zumino-gauge analogue of the Konishi
 anomaly in superspace \cite{CPS79,KO81}.

 The anomalous Slavnov-Taylor identity (\ref{STanomalous}) is the
 defining symmetry of higher-order Green functions. The anomaly part
 turns out to have indeed important implications for the
 $\beta$-functions of the gauge coupling and as we will show here, for
 the $\beta$-function of the scalar mass $M$. 

 From the ghost equations
 \begin{equation}
 \frac {\delta \Ga}{ \delta c }  = 
 \frac {\delta \Gacl}{ \delta c }\ ,  \qquad
 \frac {\delta \Ga}{ \delta \tilde c } =
 \frac {\delta \Gacl}{ \delta \tilde c }  \ ,
 \end{equation}
 one derives the gauge Ward  identity    and the anomalous Ward identity
 of axial symmetry \cite{KRST01}.

 In addition to these symmetries the Green functions are invariant under
 charge conjugation and parity.
 R-parity as defined in \cite{KRST01} is broken by the mass term of the
 photino and the left-right mass term of scalars. However,  R-parity is
 defined  by  a global $U(1)$-symmetry with a discrete transformation
 angle. The corresponding 
  global Ward identity can be derived 
  also in the case of soft breaking and includes the mass shifts
 expressing  soft breaking of the global $U(1)$ symmetry:
 \begin{equation}
 \label{WIR}
 {\cal W}^ R \Ga  = 0
 \end{equation}
 with
 \begin{eqnarray}
 \label{WIRoperator}
 {\cal W}^ R  & = &  i \intd \Bigl( \sum_{A= L,R}\bigl(\varphi_A
 \frac {\delta} {\delta \varphi_A }  -
 Y_{\varphi_A}
 \frac {\delta} {\delta Y_{\varphi_A} }\bigr) +
   {\lambda^ \alpha}
 \frac{\delta}{\delta \lambda^\alpha} -
 Y_\lambda^ \alpha
 \frac {\delta} {\delta Y_\lambda^ \alpha } \\ 
  & & \phantom{i \intd} { }  +
 {\lambdaV ^ \alpha}\frac{\delta}{\delta \lambdaV^\alpha}
 -  q ^\alpha \frac {\delta} {\delta q ^ \alpha} - \chi^ \alpha
 \frac{\delta}{\delta \chi^ 
 \alpha} \nonumber \\
 & & \phantom{i \intd} { } 
  - 2 (q_F -b)  
 \frac {\delta}{\delta q_F} 
  - 2 (f + \frac {M_\lambda} {e^2})
 \frac{\delta}{\delta f} - {\rm c.c.} \Bigr)  \nonumber  \\
 & &{ } +  i \Bigl( \epsilon^ \alpha \frac{\partial}{\partial
 \epsilon^ \alpha} - \epsilonbar^ \alphadot \frac{\partial}{\partial
 \epsilon^ \alphadot} \Bigr) \ . \nonumber
 \end{eqnarray} 
 R-parity in the usage of transforming all susy fields to their
 negative is maintained but its consequences are 
 already included in the global Ward identity
 (\ref{WIR}).

 Together with the topological formula (\ref{topfor}),
 the anomalous Slavnov--Taylor identity (\ref{STanomalous}), the
 identity 
 (\ref{holomorph}) and the global
 Ward identity (\ref{WIR})  uniquely define 
 the 1PI Green functions of the extended
 model up to invariant counterterms. Taking the  fields of the axial
 vector multiplet  and the fields of the $q$ multiplet to zero we find in
 limit to constant coupling the 1PI Green functions of softly broken
 SQED, which we denote with $ \Ga^ {\rm sSQED}$: 
 \begin{equation}
 \lim_{E \to e} \Ga \Big|_{V^ i = 0 \atop q^i= 0} =  \Ga ^ {\rm
 sSQED} \ .
 \end{equation}

\section{Symmetric counterterms}

 The appearance of symmetric counterterms
gives a hint on the divergences of the model. In particular,
absence of symmetric counterterms means that the corresponding Green
functions are determined by non-local expressions, which cannot appear
with independent divergences. In \cite{KRST01} we have found that
counterterms to the chiral vertices and counterterms to the photon
self energy in $l\geq 2$ are excluded as a consequence of
supersymmetry and the identity (\ref{holomorph}). The same arguments
apply here in the model with soft supersymmetry breaking
 and we find the same types
of symmetric counterterms as in the supersymmetric model.
Due to the shifts in the external fields, the symmetric counterterms
include now the soft breaking parameters as functions of the
supersymmetric parameters.

Symmetric counterterms are invariants with respect to the symmetries
of the model. However,
a general classical solution, i.e.~a local
 action satisfying the anomalous Slavnov--Taylor identity, does not
 exist. Thus, an invariant regularization
scheme for the extended model cannot be constructed and
  invariant counterterms can only be given order by order in the
perturbative expansion. As such,
they  are restricted by the symmetries of the
classical action:
\begin{equation}
\label{ctconstraints}
s_{\Ga_{\rm cl}} \Ga^ {(l)}_{\rm ct,inv}  =  0 \ , \qquad
\intd \Bigl(\frac {\delta } {\delta \eta}  - \frac {\delta} {\delta \etabar}
\Bigr) \Ga^ {(l)}_{\rm ct,inv}  =  0 \ , \qquad {\cal W}^ R \Ga^
{(l)}_{\rm ct,inv} = 0 \ ,
\end{equation}
and 
$\Ga_{\rm ct, inv}$ is invariant under the discrete symmetries C and
P. A further constraint on the counterterms is the topological formula
(\ref{topfor}), which determines the order in the local coupling.

Owing to these restrictions we have five types of invariant counterterms:
a  one-loop counterterm to the kinetic term of the photon multiplet,
a counterterm to the matter term of the action and three gauge
dependent 
field redefinitions for the matter fields. The counterterms are best
expressed in form of $\brs_{\Gacl}$-invariant operators acting on the
classical action. The five invariant  operators corresponding to the
invariant counterterms are given in their general form in 
appendix B. Here we discuss   the limit to constant coupling.

\begin{itemize}
\item The one-loop counterterm to the kinetic term of the photon
multiplet $\Ga^ {(1)}_{\rm ct,  kin}$ is determined by the symmetric
 operator ${\cal D}_{\rm
 kin}$ (\ref{Dkin}).
Using  the identity
\begin{equation}
\lim _{E\to e}\intd  M_{\lambda}  \Bigl(
\frac{\delta} {\delta f} + \frac{\delta} {\delta \bar f} \Bigr)
\Gacl  = e^2 M_\lambda \partial_{ M_\lambda} \Ga_{\rm cl}^ {\rm sSQED}
 \ ,
\end{equation}
it reads in the limit to constant coupling
\begin{eqnarray}
\label{Dkine}
\lim_{E\to e}\Ga^ {(1)}_{\rm ct, kin} &= &-\frac 12 e^ 2  \Bigl(
e \partial_e  + 2  {M_{\lambda}} \partial_{M_{\lambda}}
 - N_A -  N_\lambda  \\
& & \phantom{-\frac 12 e^ 2} - N_c + N_{Y_\lambda}+ N_B +
N_{\cbar}
- 2 \xi \partial_\xi \Bigr) \Ga _{\rm cl}^ {\rm sSQED}  \nonumber
\end{eqnarray}
with
\begin{eqnarray}
N_\phi &= &\intd \dfunc{ }{\phi} \ , \quad \mbox{if $\phi$ is a
real field, } \nonumber \\
N_\phi &= &\intd( \dfunc{ }{\phi} + \dfunc{ }{\bar
\phi}) \ , 
 \quad \mbox{if $\phi$ is
a complex field.}
\end{eqnarray}
Hence, the symmetric counterterm $ \Ga ^{(1)}_{\rm ct, kin} $
describes the  renormalization of the 
gauge coupling,  field redefinitions of the photon and photino and the
renormalization of the photino mass. 
\item The counterterm to the matter part of the action is decomposed
into a field redefinition of matter fields and additional invariant
counterterm $\Ga_{Vv}$. In its general form it contains redefinitions
of the axial vector multiplet into  components of the supercoupling
$E^ {2l}$ and the $q$-field renormalization. It is determined by the
operator ${\cal D}_{Vv}$ (\ref{DVv})
and results in the limit to constant coupling
in  a counterterm for the matter mass terms and the  $q$-vertices:
\begin{eqnarray}
\label{GactVv}
\lim_{E\to e} \Ga^ {(l)}_{{\rm ct},Vv} & = & - 2 e^ {2l} \Bigl(N_{ q^i} +
m\partial_ m \frac 12  l (l+1) \frac{ M_\lambda^ 2}{ M^2}  
M \partial_ M \\
& & \phantom{-2 e^ {2}} { 
} + \bigl(2l \frac {M_\lambda m}{b} +1 \Bigr) b
 \partial_
b \bigr)
 \Ga_{\rm
cl}^ {\rm sSQED} \ . \nonumber 
\end{eqnarray}
There we have used the following relations:
\begin{eqnarray}
m\partial_m \Gacl &  = &   m \intd \Bigl(\frac \delta {\delta q} +\frac\delta
{\delta \qbar  } \Bigr)\Gacl \ ,
\nonumber \\
M\partial_M \Gacl &  = & - 4 M^2 \intd \frac \delta {\delta D} \Gacl \ ,
\nonumber \\
b\partial_b \Gacl &  = &   - b \intd \Bigl(\frac
\delta {\delta q_F} + \frac \delta 
{\delta \qbar_F  } \Bigr)\Gacl\ ,
\end{eqnarray}
which make possible to eliminate for constant coupling
the field differentiation appearing
in the symmetric operator ${\cal D}_{Vv}$  
in favour of a mass
derivative.

\item There are three types of gauge dependent field
redefinitions (see (\ref{ctvarphi}),((\ref{ctpsi}) and (\ref{ctpsivarphi})).
 Two of them correspond  to the individual field
redefinitions of electrons and selectrons:
\begin{eqnarray}
\label{Gactvarphi}
\lim_{E\to e}\Ga^ {(l)}_{{\rm ct}, \varphi} = 
e^ {2l} f^ {(l)}_{\varphi} (\xi) (N_{\varphi_L} + N_{\varphi_R} 
- N_{Y_{\varphi_L}} - N_{Y_{\varphi_R}})  \Ga_{\rm cl}^ {\rm sSQED}\ , \nonumber \\
\lim_{E\to e}\Ga^ {(l)}_{{\rm ct}, \psi} = e^ {2l} f^ {(l)}_{\psi} (\xi)(N_{\psi_L} + N_{\psi_R} 
- N_{Y_{\psi_L}} - N_{Y_{\psi_R}})  \Ga_{\rm cl}^ {\rm sSQED} \ .
\end{eqnarray}
 The third one redefines an electron into a selectron and the spinor
 component of the local coupling. For constant coupling
there remains only  a  contribution  in the external field part:
\begin{equation}
\label{ctvarphipsi}
\lim_{E\to e}\Ga^ {(l)}_{{\rm ct}, \psi\varphi} = - 2 l e^ {2l}
 f_{\psi\varphi}^ {(l)}(\xi)
 \sqrt 2 M_\lambda \Bigl(
\epsilon^ \alpha Y_{\psi_L \alpha } \varphi_L \! -
\epsilonbar_\alphadot Y_{\psibar_L}^  \alphadot \phibar_L \! + (_{L\to
R})
\Bigr)  
\end{equation}
Its appearance shows that we have to expect an independent divergence
in the external field part describing the supersymmetry transformation of the
electron. Since the counterterm is linear in propagating fields, it
cannot be inserted into loop diagrams and it is not relevant for the
definition of physical Green functions.
\end{itemize}

We want to note that there is no symmetric counterterm corresponding
 to a field redefinition
of selectron field $\varphi_L$ into the right-handed selectron field
$\phibar_R$, because it is  excluded
by the global Ward identity ${\cal W}^ R \Ga_{\rm ct, inv} = 0$.

The  action of invariant counterterms in loop order $l$  is a linear
combination of 
the symmetric counterterms (\ref{Dkine}), (\ref{GactVv}),
(\ref{Gactvarphi}) and (\ref{ctvarphipsi}):
\begin{eqnarray}
\label{Gactinv}
\Ga^ {(l)}_{\rm ct, inv} & = &
 z_{\rm kin}^ {(1)} \Ga^ {(1)}_{\rm ct,  kin} \delta_{l1} +
z_{Vv}^ {(l)}\Ga^ {(l)}_{{\rm ct},Vv}  \nonumber \\
& +&
z_\varphi^ {(l)}\Ga^ {(l)}_{{\rm ct}, \varphi}(\xi) +
z_\psi^ {(l)}\Ga^ {(l)}_{{\rm ct}, \psi}(\xi) +
z_{\psi\varphi}^ {(l)}\Ga^ {(l)}_{{\rm ct}, \psi\varphi} (\xi)\ .
\end{eqnarray}

It is illuminating to rewrite these counterterms as parameter and
field renormalizations in the classical action:
\begin{equation}
\label{redefinitions}
\Gacl ((1 + z^ {(l)}_e) e, (1 + z_\phi^ {(l)}) \phi, (1 + z^ {(l)}_{M_i})M_i) =
\Gacl (e,\phi, M_i) + \Ga_{\rm ct,inv}^ {(l)} + {\cal O } (\hbar^
{2l})
\end{equation}
Comparing the $z$-factors in (\ref{redefinitions}) with the ones in
(\ref{Gactinv}) yields:
\begin{equation}
z^ {(1)}_e  = -  \frac 12 e^ 2 z^ {(1)}_{\rm kin}  \quad \mbox{and} \quad
z^ {(l)}_m   =   - 2 e^ {2l} z_{Vv}^ {(l)} \ ,
\end{equation}
 and we obtain for the $z$-factors of soft mass parameters 
\begin{eqnarray}
\label{zsoft}
%
z^ {(1)}_{M_\lambda} &= &   2  z^ {(1)}_e \ ,\nonumber \\
z^ {(l)}_b & = &  (2l \frac {M_\lambda m} b +1) z_m^ {(l)} \ ,\quad
z^ {(l)}_M =  \frac 12  l(l+1) \frac {M_\lambda^ 2} {M^ 2} z_m^ {(l)}
\ .
\end{eqnarray}  
Thus, the $z$-factors of the soft mass parameters are entirely
expressed in terms of $z_e$ and $z_m$. The same relations have to hold
for the independent symmetric divergences of the corresponding loop diagrams
when using a regularization scheme with an UV regulator.

These results are now exemplified at the one-loop level using
dimensional regularization. Although this scheme breaks supersymmetry,
the divergent one-loop contributions preserve all symmetry
constraints. 

In dimensional regularization, the one-loop divergences for the self
energies read ($\alpha = \frac {e^2} {4 \pi })$:
\begin{eqnarray}
\Sigma^ {T \rm div}_{A} & = & 
\frac{\alpha}{4\pi}2p^2 \Delta\ ,\\
\Sigma^ {\rm div}_{\lambda} & = &
\frac{\alpha}{4\pi}2p_\mu\gamma^\mu \Delta\ ,\\
\Sigma^ {\rm div}_{\varphi_L\phibar_L} & = & \frac{\alpha}{4\pi}
(-4M_\lambda^2 - 4m^2)\Delta\ ,\\
\Sigma^ {\rm div}_{\varphi_L\phibar_R} & = &
\frac{\alpha}{4\pi}(4mM_\lambda-2b)\Delta\ ,\\
\Sigma^ {\rm div}_{\Psi\bar {\Psi}} & = & \frac{\alpha}{4\pi}(2p_\mu\gamma^\mu
 -4m)\Delta \ ,
\end{eqnarray}
where $\Psi$ is the electron Dirac spinor composed of the left and
right-handed Weyl spinors $ {\psi_L}_\alpha$ and $ \psibar_R^\alphadot$.
 The divergences for $D\to4$ appear in the combination
$\Delta=\frac{2}{4-D}-\gamma_E+\log4\pi$. To absorb them, the
divergent parts of the symmetric counterterms have to be chosen as
follows:
\begin{eqnarray}
z_e^{(1)} & = & \frac{\alpha}{4\pi}\Delta  \qquad \mbox{and} \qquad
z_m^{(1)}  =  \frac{\alpha}{4\pi}2\Delta \ ,\\
z_{M_\lambda}^{(1)} & =  &  \frac{\alpha}{4\pi}2\Delta
 =2z_e^{(1)}\ ,\\
z^ {(1)}_{M} & = &  \frac{\alpha}{4\pi}2\Delta \frac{M_{\lambda^2}}{M^2}
 = z_m^{(1)}\frac{M_{\lambda^2}}{M^2}\ ,\\
z_b^{(1)} & = & \frac{\alpha}{4\pi}\Delta \left(2+
 \frac{4mM_{\lambda}}{b}\right)
 = z_m^{(1)}\left(1+\frac{2mM_{\lambda}}{b}\right)\ .
\end{eqnarray}
Indeed, these expressions are in agreement with the general results
(\ref{zsoft}).

Like in the supersymmetric model
 we find  in addition that 
independent symmetric counterterms to the chiral $q$-vertex and
to  the photon self energy in $l\geq 2$ are absent expressing the
non-renormalization theorem of chiral vertices and the generalized
non-renormalization theorem of the photon self energy.

With the same techniques as in \cite{KRST01}
we are able to relate the photon self energy in $l \geq  2 $
and the chiral vertices
to non-local expressions. The explicit expressions are modified by soft
contributions but the content and the analysis of non-renormalization
theorems is the same as in the supersymmetric case: Chiral Green
functions are up to the gauge dependent wave function renormalization
related to superficially convergent Green functions, whereas the
photon self energy is related to linearly divergent Green functions,
whose divergent part is determined  from non-local expressions via
gauge invariance.

Hence,  non-renormalization theorems and the relations of soft
parameter renormalization to the supersymmetry parameters are implied
by the same symmetries, namely by the multiplet structure of
supersymmetric Lagrangians and by the identity (\ref{holomorph})
which identifies the coupling as the lowest component of a constrained real
 superfield.

\section{Normalization conditions}

Renormalization of softly broken theories is only complete, when the
coefficients of symmetric counterterms are fixed by suitable
normalization conditions. Then symmetries and normalization conditions
together define the Green functions independently from properties of a
specific scheme used for the subtraction of divergences.

The $z$-factors $z_{\rm Vv}$, $z_{\psi}$ and $z_{\varphi}$
in (\ref{Gactinv})
appear in the same way
 as in the supersymmetric
model and 
 can be fixed for constant coupling by a normalization condition on
the electron mass and on the residua of matter fields
(see \cite{HKS99}). Similarly, the 1-loop parameter $z^ {(1)}_{\rm kin}$ 
is determined by a normalization condition on the photon residuum in
one-loop order. With  local gauge coupling
 the photon self energy in $l \geq     2$ is determined by
non-local Green functions, and  it is not possible to dispose of it
by a normalization condition without a modification of  the defining
symmetries. 


In  softly broken SQED there remains in addition a contribution from the
gauge dependent parameter
 $z_{\psi\varphi}$ (\ref{ctvarphipsi}). 
 As a normalization condition one
can require that
the respective vertex function vanishes at the normalization point
$\kappa^ 2 $:
\begin{equation}
\Ga_{\epsilon^ \alpha Y_{\psi_L}^ \beta\varphi_L} (-p,p) \Big|_{p^2 =
\kappa^2 } = 0 \ .
\end{equation}

To complete the definition of softly broken SQED 
 we have to prove that the theory as constructed here has a
  physical meaning in the sense that all fields can be interpreted as
  particles, i.e.\  it has to be shown that
 the two-point Green functions have a pole in perturbation theory.
For this purpose we  impose pole conditions on the 2-point functions
 and prove that these conditions
 are in agreement with the defining symmetries of
 the model. 

Apparently, there are   not enough  symmetric 
counterterms for setting 
normalization conditions of  soft parameters, but these  can be imposed
when we include
 finite redefinitions of the
mass parameters. In the extended model with local coupling such
redefinitions
induce 
higher order corrections to the shifts in the Slavnov-Taylor identity
and in the Ward-identity of R-symmetry (\ref{WIR}):
\begin{eqnarray}
\label{softonshell}
f(x) & \to & f(x) + \frac 1 {e^ 2} ({M_\lambda} + \sum _{l=1} ^ \infty v^
{(l)}_\lambda e^ {2l})\ , \nonumber \\
q_F & \to & q_F(x) - ( b   + \sum_{l=1}^ \infty v_b^
{(l)} e^ {2l})\ , \nonumber  \\
\DV & \to & \DV  - 2 ( M + \sum_{l=1}^ \infty v^ {(l)}_M e^ {2l})^2
\ ,
\end{eqnarray} 
but do not change the structure of symmetry transformations.

In  this context we want 
to note that due to parity conservation
 mass eigenstates of the selectron fields are
constructed
to all orders  by an orthogonal transformation on the left- and right-handed
selectron fields 
\begin{eqnarray}
\label{mattereigenstates}
\varphi_1 = \frac 12 \sqrt 2 (\varphi_L  - \phibar_R)\ , & \qquad &
\phibar_1 = \frac 12 \sqrt 2 (\phibar_L  - \varphi_R) \ ,
 \nonumber \\
\varphi_2 =  \frac 12 \sqrt 2 (\varphi_L  + \phibar_R)\ , & \qquad &
\phibar_2 = \frac 12 \sqrt 2 (\phibar_L  + \varphi_R) \ .
\end{eqnarray}
These fields are also eigenstates with respect to charge conjugation
and parity transformation:
\begin{equation}
\varphi^ C_1  = - \phibar_1\ ,  \quad \varphi^ C_2  =  \phibar_2 \quad
\mbox{and} \quad
\varphi^P _1  = - \phibar_1\ ,  \quad \varphi^ P_2  =  \phibar_2 \ ,
\end{equation}
and the vertex function $\Ga_{\varphi_2 \phibar_1} $ vanishes due to
parity conservation. 
With the finite redefinitions (\ref{softonshell})
 pole conditions for the photino, and the scalar fields
$\varphi_1$ and $\varphi_2 $ can be established by adjusting the parameter
$v_\lambda$, $v_b$ and $v_M$ as functions of the mass parameters of
softly broken SQED.

The appearance of
the higher order shifts  (\ref{softonshell}) does not change the
analysis of symmetric counterterms and the content and analysis of
non-renormalization theorems. However,  the
Callan--Symanzik and renormalization group functions which we
determine
 in the subsequent sections are scheme-dependent and
 depend on the specific form of the normalization conditions and
symmetries.
 It is possible to include the higher-order
shifts induced by pole conditions of soft parameters 
 into the construction without difficulty, but
 their appearance obscures the
algebraic structure of the Callan-Symanzik and renormalization group
coefficients. For this reason we will restrict  to the classical
shifts for the remaining part of the paper and note that these
conditions
match
 the ones used in the MS-scheme of dimensional reduction.
For generalized normalization conditions, the respective
expressions
 can be obtained by  carrying out a redefinition of the
form (\ref{softonshell}) in the Callan--Symanzik equation
(\ref{CS}) and the
related mass equations  (\ref{massequ}).


\section{The Callan--Symanzik equation}

The relations of soft breaking parameters to the supersymmetric
parameters as well as the non-renormalization theorems have immediate
implications for the Callan--Symanzik coefficients and renormalization
group coefficients of softly broken SQED. Indeed, from the symmetries
of the model with local gauge coupling and gauged axial symmetry we
obtain the gauge $\beta$-function in its closed form and the all-order
expressions for the  anomalous mass dimensions and
$\beta$-functions of soft parameters.
 We start the construction
with the Callan--Symanzik equation and continue the analysis to the
renormalization group equation in the next section. 

The Callan--Symanzik (CS) equation is the partial differential equation
connected  with the breaking of dilatations. The dilatations act on 
the Green functions in the same way as a scaling of all mass
parameters of the theory including the normalization point $\kappa$
according to their mass dimension: 
\begin{equation}
{\cal W}^ D \Ga = 
- ( m\partial_m + M_\lambda \partial_{M_\lambda}
+ M \partial_M + 2 b \partial _b +\kappa \partial_\kappa) \Ga 
\equiv 
-\mu\partial_{\mu}  \ .
\end{equation}

At the tree level, dilatations are broken  by the  vertex functions
 with non-vanishing mass
dimension. 
 By means of the external fields we obtain the following
 expression in the classical approximation:
\begin{eqnarray}
\label{CSclassical}
\mu \partial_{\mu} \Gacl & =
& m \intd \Bigl(\frac {\delta} {\delta q} + \frac {\delta} {\delta
\qbar}\Bigr) \Gacl +  {M_\lambda}
  \intd \frac 1{e^ 2}
\Bigl(\frac {\delta} {\delta f} + \frac {\delta} {\delta
\fbar}\Bigr) \Gacl  \nonumber \\
& & -4  M^ 2
  \intd \frac {\delta} {\delta \DV}  \Gacl
- 2 b  \intd \Bigl(\frac {\delta} {\delta q_F} + \frac {\delta} {\delta
\qbar_F}\Bigr) \Gacl 
\ .
\end{eqnarray}
We rewrite the classical equation (\ref{CSclassical}) in the form
\begin{equation}
\mu D_{\mu} \Gacl = 0\ ,
\end{equation}
and note that $\mu D_{\mu}$ is a symmetric operator with respect to the
anomalous Slavnov--Taylor operator (\ref{STanomalous}) and with respect to the
Ward  operator of softly broken $R$-symmetry (\ref{WIR}).

In higher orders the CS equation of softly broken SQED
is constructed in the same way as  for SQED. For this reason we skip  the
construction and refer for details to  \cite{KRST01}.
There it was shown that the CS operator has to be
constructed as a symmetric operator with respect to the anomalous
Slavnov--Taylor identity. We find  five independent $\brs_\Gacl$-invariant
operators with the correct quantum numbers -- they are just the ones
 which we have already used for the construction of
invariant counterterms. These five operators
can be extended to $\brs_\Ga + r^ {(1)} \delta
{\cal S}$-symmetric operators ${\cal D}_{\rm kin},{\cal D}^ {{\rm sym}
}_{Vv},
{\cal N}_\varphi,{\cal N}_\psi$ and ${\cal
N}_{\psi\varphi}$.
  The CS operator is composed as a linear combination of these five operators:
\begin{eqnarray}
\label{CSop}
{\cal C}& = & \mu_i D _{\mu_i} 
+ \hat \beta_e^ {(1)} {\cal D}_{\rm kin}
 - \sum_{l}\bigl(\hat \gamma ^ {(l)} {\cal D}^ {{\rm sym} (l)}_{Vv} +
\hat \gamma_\varphi^ {(l)} {\cal N}_\varphi^ {(l)}
+ \hat \gamma_\psi^ {(l)} {\cal N}_\psi^ {(l)} +
\hat \gamma_{\psi\varphi}^ {(l)} {\cal N}_{\psi\varphi}^ {(l)}
 \bigr).
\end{eqnarray}
 The algebraic construction
yields the CS equation 
\begin{equation}
\label{CS}
{\cal C}\Ga = \Delta_Y 
\end{equation}
for the extended model of SQED with soft breakings. In (\ref{CS})
$\Delta_Y$ 
is a field monomial linear in propagating fields and contains
those parts of invariants, which are not expressed in form of
operators. It is also uniquely defined in the algebraic construction
and depends on the CS coefficients of the left-hand side.

The invariant operators appearing in the 
CS operator (\ref{CSop}) are given in their general form in 
appendix  B,  here we want to point out the property that the symmetric
operator
${\cal D}_{Vv}
^{{\rm sym}(l)}$ includes not only an $l$-loop operator,
but also operators of order $l+1$:
\begin{equation}
\label{DvVsym}
{\cal D}^ {{\rm sym}(l)}_{Vv} \equiv
{\cal D}^ {(l)}_{Vv} - r^ {(1)}\bigl( 4 {\cal D}^ {(l+1)}_{e} +
8 l ( {\cal N}^ {(l+1)}_V - 8 (l+1)r ^{(1)} \delta {\cal N}^ {(l+2)}_V )
 \bigr) .
\end{equation}
In this expression the operators of  order $l+1$   are determined by the
anomaly absorbing part  $r^ {(1)} \delta
{\cal S} $ of the anomalous Slavnov--Taylor identity.
 The  operator ${\cal D}_e$ contains a differential operator with
respect to the gauge coupling and determines
 the gauge $\beta$-function in its closed form   \cite{CPS79, VZS85}.
 The operator ${\cal N} _V$ describes an
anomalous dimension of the axial vector multiplet and generates
the $X$-term to the scalar mass $\beta$-function \cite{JAJO98scalar2}.

In
its structure the CS
equation (\ref{CS})  coincides  with the one of the supersymmetric
model. It contains the two gauge independent coefficients
$\beta^ {(1)}$ and $\gamma$ and the gauge dependent anomalous
dimensions of matter fields, which are specific for the Wess-Zumino gauge.
Furthermore, 
the one-loop coefficients
 are mass 
independent, and are therefore   the same functions as in SQED.
In particular one has
\begin{equation}
\beta^ {(1)}_e =  e^2 \frac 1 {8\pi^2 } \quad \mbox{and} \quad
\gamma^ {(1)} = -  \frac{ e^2} {2\pi^ 2}  \ .
\end{equation}
For $l\geq 2$ the anomalous dimension $\gamma$ is in general mass
dependent and depends on the specific normalization conditions for
defining
the coefficient of the invariant counterterm  $z_{Vv}$.

 From the expression $(\ref{CS})$ we can extract the anomalous mass
dimensions of soft parameters in their usual form. For this purpose
 we set the axial vector field,  its superpartners  and
the fields of the ${ q}$-multiplet to zero. Then we find  
the  CS equation for the Green function of SQED with soft
supersymmetry breaking and constant coupling:
\begin{eqnarray}
\label{CSconst}
& & \Bigl(\mu_i\partial_{\mu _i} +
e^2(\hat \beta_e^ {(1)} + 4 r^ {(1)} \gamma) (e\partial_e - N_A - N_\lambda + N_{Y_\lambda} - N_c + N_B +
N_\cbar  - 2 \xi \partial_\xi)\nonumber \\
& & -  \gamma _\varphi (N_{\varphi_L}\! + N_{\varphi_R}\! - 
N_{Y_{\varphi_L}}\! - N_{Y_{\varphi_R}}) 
-  \gamma _\psi (N_{\psi_L} \!+ N_{\psi_R}\! - 
N_{Y_{\psi_L}}\! - N_{Y_{\psi_R}})  \Bigr)\,   \Ga^ {\rm sSQED} \nonumber \\
& &  = \biggl(
m (1-2\gamma)  \intd  \bigl(\frac{\delta}{\delta q} +
\frac{\delta}{\delta \qbar}
\bigr)  + 
  \frac {M_\lambda} {e^ 2} (1 - \gamma_{M_\lambda})
  \intd \Bigl(\frac {\delta } {\delta f} + \frac {\delta} {\delta
\fbar}\Bigr)   \nonumber \\
& & -4  M^ 2 (1 - \gamma_{M})
  \intd \frac {\delta} {\delta \DV}  
- 2 b (1 - \gamma_b)
  \intd \Bigl(\frac {\delta} {\delta q_F} + \frac {\delta} {\delta
\qbar_F}\Bigr) \biggr) \Ga \Big|_{V^ i, q^ i= 0
 \atop E \to e \hfill} \nonumber \\
& & - e\partial_e\gamma_{\psi \varphi}
 \sqrt 2 M_\lambda \Bigl(
 Y_{\psi_L}^ { \alpha } \epsilon_ \alpha\varphi_L -
 Y_{\psibar_L  \alphadot} \epsilonbar^\alphadot \phibar_L + (_{L\to
R})\Bigr) \ .
\end{eqnarray}
As usually it is an inhomogeneous partial differential equation
containing on the right-hand-side the soft mass insertions of the
breaking of dilatations.

In the CS equation (\ref{CSconst}) the anomalous dimensions of soft parameter,
$\gamma_{M_\lambda}, \gamma_{M}$ and $\gamma_b$, are entirely
determined by the anomalous dimensions of the supersymmetric mass
parameter $\gamma$ and the one-loop $\beta$-function $\beta^ {(1)}$.

Like the higher-order contributions to the gauge $\beta$-function the
anomalous dimension of the photino  mass is determined by
the operator 
 ${\cal D}_{\rm kin}$ and the
operator ${\cal D}^ {(l)}_e $:
\begin{eqnarray}
 & & \lim _{E\to e} (\hat \beta_e^ {(1)} {\cal D}_{\rm kin}+4 r^ {(1)} \sum_{l}
\hat \gamma^ {(l)} {\cal D}^ {(l+1)}_e ) \Ga  \nonumber \\
  & = &
 \beta_e  (e\partial _e +\cdots) \Ga_{E\to e}
 +  \frac {M_\lambda} {e^2} \gamma_{M_\lambda}
\intd \Bigl(\frac {\delta} {\delta f} + \frac {\delta}{\delta
 \fbar}\Bigr) \Ga\Big|_{E\to e}   \ .
\end{eqnarray}
From this expression
we obtain  the gauge $\beta$-function $\beta_e$ in its closed form,
\begin{equation}
\label{betae}
\beta_e = e^ 2 (\hat \beta^ {(1)}_e + 4 r^{(1)} \gamma) \quad
\mbox{with} \quad \gamma = \sum_l \hat \gamma ^ {(l)} e ^{2l} \ ,
\end{equation}
and the anomalous dimension $ \gamma_{M_ \lambda}$,
\begin{equation}
\label{gammaMlambda}
\gamma_{M_\lambda} =
\bigl(2 e ^2 (\hat \beta_e^ {(1)} + 4 r^ {(1)}  \gamma)  + 8 r^ {(1)}
\sum_l \hat \gamma^ {(l)} l e ^{2(l+1)}\bigl) \ ,
\end{equation}
  in terms of $\hat\beta_e^ {(1)}$, $\hat \gamma^
{(l)}$
and the anomaly coefficient
$r^ {(1)}$.

 Contributions to the anomalous mass dimension of the
 scalar mass $M$ arise from the operator
${\cal D}_{Vv}$ and from ${\cal N}_V$. In addition,
from ${\cal D}_{Vv}$  one obtains 
 the 
anomalous  dimension of the $b$ parameter.
Evaluating the operators ${\cal D}_{Vv}$ (\ref{DVv})  and ${\cal N}_V$ 
(\ref{NV}) for constant coupling
 we find:
\begin{eqnarray}
\label{gammaM}
\hat \gamma_{M}^ {(l)} & = & \hat\gamma^ {(l)}
(  l (l+1) \frac {M_\lambda^ 2} {M^2} + 4 e^ 2 r^ {(1)} l    )  \ ,\\
\label{gammab}
\hat \gamma_b^{(l)}
& = &\hat\gamma^ {(l)}\bigl(1 +  2 l \frac {M_\lambda m} b) \ .
\end{eqnarray}

Finally it is possible to rearrange  the expressions
for $\gamma_{M_\lambda}$, $\gamma_{b}$ and $\gamma_{M}$ into a more
familiar form by applying
the
differential operator $e\partial_e$  on the
gauge $\beta$-function $\beta_e$ and the anomalous dimension
$\gamma$ (\ref{betae}). 
Defining the anomalous dimensions of soft mass parameters as a power
series in the coupling
\begin{equation}
\gamma_{ M} = \sum_l e ^{2l} \hat \gamma_{M} ^ {(l)} 
 \quad \mbox{and}  \quad \gamma_b = \sum_l e ^{2l} \hat \gamma_b ^
 {(l)}\ ,
\end{equation}
one obtains from (\ref{gammaMlambda}), (\ref{gammaM}) and
(\ref{gammab}) the following expressions:
\begin{eqnarray}
\label{gammasoft1}
\gamma_{M_\lambda}  & = &  e \partial_e ( \beta^ {(1)}_e + 4 r^ {(1)} e^2
\gamma )    \ ,\\
\label{gammasoft2}
\gamma_{M} & = & \frac  { M_\lambda^ 2}
{4M^2} e\partial_e \bigl (e \partial_e \gamma) + \frac
{ M_\lambda^ 2 }{2M^ 2} e\partial_ e \gamma + 2 r^ {(1)}  e^ 3
\partial_e \gamma  \ ,\\
\label{gammasoft3}
\gamma_b & = &  \gamma  + \frac { M_\lambda m}b e \partial_e \gamma \ .
\end{eqnarray}
These expressions coincide
with similar expressions given for the $\beta$-functions of soft mass
parameters in refs.~\cite{JAJO97gaugino, AKK98} including  the $X$-term
for the scalar mass 
$\beta$-function  \cite{JAJO98scalar2}. The relations of the anomalous
mass dimensions in the CS equation to 
the renormalization group $\beta$-functions  
are discussed finally in section 7. 

In summary we find that
the CS equation of softly broken SQED contains  the
same gauge independent coefficient as the CS equation of
SQED. These are the one-loop $\beta$-function and the anomalous mass
dimension
$\gamma$ appearing with the renormalization of the supersymmetric mass
parameter  $m$. Hence, the gauge
$\beta$-function is expressed in its closed form,  and
one obtains a common anomalous dimension for the chiral supersymmetric
vertices.
 Both are implications of the non-renormalization theorems
for  chiral vertices and for
the photon self energy in $l \geq 2$.  Moreover, and this is the 
remarkable point of the construction, 
the anomalous contributions to the soft breakings are completely
expressed in terms of the two  gauge independent CS
coefficients $\beta_e^{(1)}$, $ \gamma$ and the anomaly coefficient
$r^{(1)}$.

\section{The renormalization group equation and the 
limit to supersymmetric QED}

The CS equation is not the only partial differential
equation  for scaling of mass parameters in softly broken SQED, but
 we can derive similar
equations as the CS equation for the differentiation with
respect to all mass parameters of the theory.

The corresponding mass equations 
 take the general form ($M_i =  M_\lambda, M, m, b$):
\begin{eqnarray}
\label{massequ}
& &\bigl( M_i D_{ M_i} 
+ \hat \beta_e^{i (1)} {\cal D}_{\rm kin}  \nonumber \\
& &  - \sum_{l}(\hat \gamma ^ {i (l)} {\cal D}^ {{\rm sym}
(l)}_{Vv} +  \hat \gamma_\varphi^ {i (l)} {\cal N}_\varphi^ {(l)} 
+ \hat \gamma_\psi^ {i(l)} {\cal N}_\psi^ {(l)} +
\hat \gamma_{\psi\varphi}^ {i (l)} {\cal N}_{\psi\varphi}^ {(l)}
 )\bigr) \Ga  =  \Delta_Y^ {i} \ .
\end{eqnarray}
Here $M_iD_{M_i}$ are the symmetric operators corresponding to the mass
differentiation $M_i \partial_{M_i}$. Explicitly one has:
\begin{eqnarray}
mD_m &\equiv & m\partial_m - 
 m \intd \Bigl(\frac {\delta} {\delta q} + \frac {\delta} {\delta
\qbar}\Bigr)  \ ,\\
M_\lambda D_{M_\lambda} &\equiv& M_{\lambda} \partial_{M_\lambda} -
  \intd \frac {M_\lambda}{e^ 2}\Bigl(\frac {\delta} {\delta f} + \frac {\delta} {\delta
\fbar}\Bigr) \Ga \ , \nonumber \\
M_\lambda D_{M_\lambda} &\equiv&M\partial_M \Ga + 4  M^ 2
  \intd \frac {\delta} {\delta \DV}  \Ga\ , \nonumber \\
b D_b &\equiv&b \partial_b 
+ b  \intd \Bigl(\frac {\delta} {\delta q_F} + \frac {\delta} {\delta
\qbar_F}\Bigr) \Ga \ . \nonumber
\end{eqnarray}
 Starting from the classical equations it is obvious, that the
breaking  of higher orders in eq.~(\ref{massequ})
is  a linear combination of
 the  five $\brs_\Ga + r^{(1)} \delta {\cal S}$-invariant operators,
which we had used in the 
construction of the CS
equation.
 Hence,
in the limit to constant coupling we find among the coefficient
functions of the individual mass equations the same relations 
 as for the coefficient functions of the 
CS equation (see (\ref{betae}) and  (\ref{gammasoft1}) 
-- (\ref{gammasoft3})).
The $\beta$ function $\hat \beta^{i(1)}_e $ and the anomalous
dimension $
\hat \gamma^ {i(l)}$ appearing
in (\ref{massequ})  are scheme dependent and
depend on the specific normalization
conditions already in one-loop order.

By subtraction of the single mass equations from the CS
equation we obtain  the renormalization group (RG) equation for the
variation of the normalization point $\kappa$. 
 The $\beta$-functions of the RG equation can be
compared with the $\beta$-functions of soft mass parameters as defined
in refs.~\cite{JAJO97gaugino, JAJO98scalar2, AKK98}.
 The latter are determined in specific mass
independent schemes.
 The intrinsic normalization conditions of
such schemes are normalization conditions at an asymptotic
normalization point 
$\kappa_\infty^ 2 $  which is  much larger than all mass
parameters  of
the theory \cite{KR94}:
\begin{equation}
|\kappa_\infty ^ 2| \gg m^2, M^ 2_\lambda, b,M^2 \ .
\end{equation} 
We take the usual normalization conditions for 
 the residua of matter fields and of the photon
field in one-loop at the asymptotic normalization point
 $\kappa_\infty^ 2$ \cite{HKS99}.
For  a mass independent definition of the  invariant
counterterm $\Ga_{{\rm ct}, Vv}$  (\ref{GactVv})
we choose a normalization condition  on a dimensionless
vertex function at the symmetric point as for example:  
\begin{equation}
\label{normvVas}
\Ga_{q  \psi_{R \alpha} \psi_L^ \alpha} \Big|_{p_i^ 2 =  \kappa_\infty^2 
\hfill\atop
 2 p_i p_j = -  \kappa_\infty^ 2, i \neq j} = 2 \ .
\end{equation}

Evaluating the equations (\ref{massequ}) for the normalization
vertices at the asymptotic normalization point
 we find
that the mass equations 
are in their tree form to all orders of perturbation theory, i.e.
\begin{equation}
\label{massasymptotic}
M_i D_{M_i}\Ga = 0\ , \qquad M_i = m, M, M_\lambda, b \ .
\end{equation}
(We omit the contributions  $\gamma_{\psi\varphi}$  of the trivial
insertion in the subsequent discussion.)

By means of  these mass equations 
 the field
differentiation on the right-hand-side of the CS equation (\ref{CSconst})
can be
 eliminated for constant coupling in favour of
a mass differentiation. The resulting equation
is  the RG
equation  of softly broken
SQED for asymptotic normalization conditions. It  is a homogeneous
 partial differential equation 
for the vertex functional of softly broken SQED $\Ga^ {\rm sSQED}$:
\begin{eqnarray}
\label{RGconst}
& & \Bigl(\kappa_\infty \partial_{\kappa_\infty} +
 \beta_e (e\partial_e - N_A - N_\lambda + N_{Y_\lambda} - N_c + N_B +
N_\cbar  - 2 \xi \partial_\xi)\nonumber \\
& &
+ \gamma_{M_\lambda} {M_\lambda}\partial_{M_\lambda}
+ \gamma_{M} {M}\partial_{M}  + 2 \gamma_b b
\partial_b
+2 \gamma  {m}\partial_{m} -  \gamma _\varphi (N_{\varphi_L} +
 N_{\varphi_R}   \nonumber \\
& & 
\quad - N_{Y_{\varphi_L}} - N_{Y_{\varphi_R}}) 
-  \gamma _\psi (N_{\psi_L} + N_{\psi_R} - 
N_{Y_{\psi_L}} - N_{Y_{\psi_R}})  \Bigr)\,   \Ga^ {\rm sSQED}  =  0 
\end{eqnarray}
Here $\beta_e $ and $\gamma_{M},\gamma_{M_\lambda}$ and $\gamma_b$ are
functions of  
$\hat \beta_e^ {(1)}$ and $\gamma$ as given in (\ref{betae}) and
(\ref{gammasoft1}) -- (\ref{gammasoft3}).
Interpreting the coefficients of mass differentiation as
$\beta$-functions for soft parameters one gets:
\begin{eqnarray}
\beta_{M_\lambda}  & = & M_\lambda  e \partial_e  {\beta_e}\ ,  \\
\beta_{M} & = & M \Bigl(\frac  { M_\lambda^ 2}
{4M^2} e\partial_e \bigl (e \partial_e \gamma) + \frac
{ M_\lambda^ 2 }{2M^ 2} e\partial_ e \gamma + \frac 12  e^ 3 \partial_e \frac
{\beta_e} {e ^2} \Bigr) \ ,\\
\beta_b & = & 2b ( \gamma  + \frac { M_\lambda m}b e \partial_e \gamma )
\ .
\end{eqnarray}
These expressions can be immediately compared with previous results on
the $\beta$-functions, finding correspondence for the
$\beta$-functions of the photino and the $b$-parameter
\cite{AKK98,JAJO97gaugino}, and the
correct all order expression for the $\beta$-function of the scalar
mass as suggested in \cite{JAJO98scalar2}.

Finally using consistency conditions of the mass equations
 (\ref{massasymptotic})  with the
 RG equation (\ref{RGconst}) we find that the coefficients
$\gamma, \gamma_\varphi$ and $\gamma_\psi$
 are
 mass independent \cite{KR94}.  For this reason  the CS
coefficients of the supersymmetric theory and the CS
 coefficients as
 well as the RG coefficients of
 softly broken SQED coincide for asymptotic normalization conditions.
 In this sense the
RG functions of the softly broken theory are indeed
determined  by the ones of  supersymmetric QED.

\section{Conclusions}
 
In the present paper the specific renormalization properties of softly
broken SQED
 have been worked out on the basis of algebraic
renormalization. The construction is based on  the
extended model of SQED with a local gauge coupling and gauged axial
symmetry. In the extended model  softly broken supersymmetry
 is a natural generalization of unbroken supersymmetry.
 By a constant shift in the highest components of external field
multiplets the  soft mass terms are generated without changing
 the structure of
defining symmetry transformations.

The non-renormalization theorems of chiral vertices and  the 
generalized non-renormalization theorem of the photon
self energy in $l\geq 2$ are deduced in the same way as in the
supersymmetric model: Chiral vertices are superficially convergent up
to gauge dependent field redefinitions, and the photon self energy can
be related to non-local expressions via gauge invariance.
Furthermore, it is seen that the symmetric counterterms of
soft mass parameters are related to the counterterms of the
supersymmetric mass  and to the  one-loop counterterm of
the photon self energy.  

These relations imply also restrictions on the $\beta$-functions of
soft mass parameters inferring the closed form of the gauge $\beta$
function and the exact all-order formulas for the  soft 
$\beta$-functions from the algebraic construction.

In comparison to related constructions (see \cite{AKK98,KO00} and
\cite{AGLR98}) the relevant difference  concerns the inclusion of
anomalous axial symmetry with the
axial vector multiplet    in
the present construction. In ref.~\cite{KRST01} we have shown, that
the axial vector multiplet has to be introduced in order to complete
SQED with the local gauge coupling to a multiplicatively
renormalizable theory. Including axial symmetry we have to care about the
Adler--Bardeen anomaly with its supersymmetric extension.
Then, of course,  a
consistent higher-order construction cannot be performed with
arguments on invariant schemes and,  in particular, 
 it is not possible to infer the RG
$\beta$-functions  from 
 symmetric counterterms in general.

It is the remarkable point in the construction that the local coupling
makes possible to absorb
the anomaly  into a modified anomalous
Slavnov--Taylor identity, and in
this case 
algebraic renormalization can be performed in presence of the anomaly.
The closed form of the gauge $\beta$-function and the all-order
expressions of soft $\beta$-functions
 are the result of the algebraic construction of the CS and RG
equation in presence of the anomaly in
the  Slavnov--Taylor identity. 

Finally we  want to remark that the present construction is not
restricted to the Wess-Zumino gauge, but can be performed in the same
way with linear supersymmetry transformations and a supersymmetric
gauge fixing
using the superspace formalism.

\vspace{0.5cm}
{\bf Acknowledgments}

We thank R. Flume and W. Hollik  for  discussions. 
E.K. is grateful to the Fachbereich Physik of the Universit\"at
Kaiserslautern for kind hospitality where  parts of this work have been
  done.

\newpage

\begin{appendix}
\section{The BRS transformations and the Slavnov-Taylor identity} 
 
In this appendix we list the BRS transformations of external fields
 including the shifts. The BRS transformations of the remaining  fields
take the conventional form with local gauge coupling and are
 not modified by the shifts. Their explicit form  can be found in
 ref.~\cite{KRST01}. 
\begin{itemize}
\item BRS transformations of the axial vector multiplet and of axial
ghost
\begin{eqnarray}
sV_\mu & = & \partial_\mu \tilde c + i\epsilon\sigma_\mu{\bar {\tilde
\lambda}}
             -i \tilde
\lambda\sigma_\mu\epsilonbar -i\omega^\nu\partial_\nu V_\mu
\ ,\\
s\tilde\lambda^\alpha & = & \frac{i}{2} (\epsilon\sigma^{\rho\sigma})^\alpha
             F_{\rho\sigma}( V) + \frac i2 \epsilon^\alpha\,
             (\tilde D - 2 M^2) -i\omega^\nu\partial_\nu  
             \tilde \lambda^\alpha
\ ,\nonumber \\
s{\bar {\tilde \lambda}}
_\alphadot & = & \frac{-i}{2} (\epsilonbar\sigmabar^{\rho\sigma})
             _\alphadot F_{\rho\sigma}(V) + \frac i 2 \epsilonbar_\alphadot 
(\tilde D - 2 M^2)
             -i\omega^\nu\partial_\nu \bar {\tilde \lambda}_\alphadot
\ , \nonumber \\
s \tilde D & = & 2 \epsilon \sigma^ \mu \partial_\mu {\overline {\tilde \lambda}} + 
                  2 \partial_
 \mu  {\tilde \lambda} \sigma^ \mu \epsilonbar 
             -i\omega^\nu\partial_\nu  {\tilde D} \ ,\nonumber \\
s\tilde  c & = & 2i\epsilon\sigma^\nu\epsilonbar V_\nu
-i\omega^\nu\partial_\nu\tilde  c
\ \nonumber 
\end{eqnarray}
\item BRS transformations of the local coupling and its superpartners
 (\ref{E2def})
\begin{eqnarray}
s \eta & = & \epsilon^ \alpha \chi_\alpha -i \omega ^ \nu \partial_\nu \eta \ ,\\
s \etabar & = & \chibar_\alphadot \epsilonbar^ \alphadot
 -i \omega ^ \nu \partial_\nu
\etabar \ ,\nonumber \\
s \chi_\alpha& = & 2 i (\sigma^ \mu \epsilonbar)_\alpha \partial_\mu \eta + 2
\epsilon_\alpha (f + \frac {M_\lambda} {e^2 }) 
- i \omega^ \mu \partial_\mu
\chi_\alpha\nonumber \\
s \chibar_\alphadot& = & 2 i (\epsilon \sigma^ \mu)_\alphadot
 \partial_\mu \etabar   - 2
\epsilonbar_\alphadot (\bar f + \frac {M_\lambda} {e^2 })
- i \omega^ \mu \partial_\mu
\chibar_\alphadot \nonumber \\
s f & = & - M (\epsilon \chi +\chibar \epsilonbar)+  i 
\partial_\mu \chi \sigma^ \mu \epsilonbar - i \omega^ \mu \partial_\mu
f \nonumber \\
s \bar f & = & - M (\epsilon \chi +  \chibar \epsilonbar) - i 
 \epsilon \sigma^ \mu \partial_\mu \chibar 
- i \omega^ \mu \partial_\mu
\bar f  \nonumber
\end{eqnarray}
\item BRS transformations of $q$-multiplets (\ref{qdef})
\begin{eqnarray}
s q & = &  + 2i \tilde c (q  + m ) +
 \epsilon^ \alpha q_\alpha -i \omega ^ \nu \partial_\nu q\ ,\\
s \qbar & = & -2i \tilde c (\qbar + m)
+ \qbar_\alphadot \epsilonbar^ \alphadot
 -i \omega ^ \nu \partial_\nu
\qbar \ , \nonumber \\
s q_\alpha& = & + 2i \tilde c q_\alpha +2 i (\sigma^ \mu \epsilonbar)_\alpha 
D_\mu q  + 2
\epsilon_\alpha (q_F  -b )
- i \omega^ \mu \partial_\mu
q_\alpha \ ,\nonumber \\
s \qbar_\alphadot& = &  - 2i \tilde c \qbar_\alphadot +
2 i (\epsilon \sigma^ \mu)_\alphadot
 D_\mu \qbar     - 2
\epsilonbar_\alphadot (\qbar_F -b)
- i \omega^ \mu \partial_\mu
\qbar_\alphadot \ ,\nonumber \\
s q_F & = &+ 2i \tilde c (q_F -b) +
 i 
D_\mu q^ \alpha \sigma^ \mu_{\alpha \alphadot} \epsilonbar^ \alphadot
- 4 i
 \lambdaVbar _ \alphadot \epsilonbar^ \alphadot (q +m )
 - i \omega^ \mu \partial_\mu
q_F \ , \nonumber \\
s \qbar_F & = & - 2i \tilde c (\qbar_F -b )  - i 
 \epsilon^ \alpha \sigma^ \mu_{\alpha \alphadot} D_\mu \qbar^ \alphadot 
+ 4 i
\epsilon ^ \alpha \lambdaV _ \alpha  (\qbar + m)
- i \omega^ \mu \partial_\mu
\qbar_ F \ . \nonumber
\end{eqnarray}
The covariant derivative is defined by
\begin{equation}
\label{covq}
D_\mu { q^ i} = (\partial _\mu - 2 i V_ \mu)({ q^ i} + (m,0,-b))
\end{equation}
\end{itemize}

The classical Slavnov--Taylor identity (\ref{ST})
expresses in functional form BRS invariance of the classical action
and on-shell  nilpotency of BRS transformations.  
 The Slavnov--Taylor operator acting on a general
functional ${\cal F}$  is defined as
\begin{eqnarray}
{\cal S}({\cal F}) & = & 
\intd\Bigl(s A^\mu \dF{A^\mu}
+ \dF{Y_\lambda{}_\alpha}\dF{\lambda^\alpha}
+ \dF{Y_\lambdabar^\alphadot}\dF{\lambdabar_\alphadot}
 \nonumber\\&&{}\quad
+   s c \dF{c}         + s B \dF{B} + s\bar{c} \dF{\bar{c}} + s \xi \dF{\xi}
+  s\chi_\xi \dF{\chi_\xi} 
\nonumber\\&&{}\quad
+ \dF{Y_{\varphi_L}}\dF{\varphi_L}
+ \dF{Y_{\phibar_L}}\dF{\phibar_L}
+ \dF{Y_{\psi_L{}_\alpha}}\dF{\psi_L^\alpha}
+ \dF{Y_{\psibar_L}^\alphadot}\dF{\psibar_L{}_\alphadot}
+(_{L\to R})
\nonumber\\&&{} \quad
+ s\eta^i\frac{\delta{\cal F}}{\delta\eta^ i}
+ s\etabar^i\frac{\delta{\cal F}}{\delta\etabar^ i}
+ s q^i\frac{\delta{\cal F}}{\delta q^ i}
+ s\qbar^i\frac{\delta{\cal F}}{\delta\qbar^ i}
\nonumber\\&&{} \quad
+ s V^i\frac{\delta{\cal F}}{\delta V^ i}
+ s \tilde c\frac{\delta{\cal F}}{\delta \tilde c}
 \Bigr)
+ s\omega^\nu \frac{\partial{\cal F}}{\partial\omega^\nu} \ .
\label{STOperator}
\end{eqnarray}

\section{Invariant operators}

The invariant operators 
 ${\cal O}^
{\rm sym}$ are defined as being symmetric with respect to
the symmetries of functional of 1PI Green functions:
They are invariant  with
respect to the anomalous Slavnov-Taylor identity:
\begin{equation}
\label{CSSTcons}
\bigl( s_{\Ga} + r ^{(1)} \delta {\cal S}\bigr) {\cal O}^
{\rm sym}   \Ga = 
{\cal O}^
{\rm sym}\bigl({\cal S}+ r^ {(1)}\delta {\cal S}\bigr) \Ga +
\bigl(s_{\Ga} + r ^{(1)} \delta {\cal S}\bigr) \Delta_Y \ .
\end{equation}
The expression $\Delta_Y $  is defined to be a collection of field monomials which are
 linear in propagating fields.
And they commute with the Ward identity of $R$-symmetry:
\begin{equation}
\big[{\cal O}^
{\rm sym} , {\cal W}^ R \Big] = 0\ .
\end{equation}
According to the identity (\ref{holomorph}) they depend on $\eta -
\etabar$ only by a derivative.
The numbers of local couplings contributing in the operator of loop
order $l$  is restricted by the topological formula (\ref{topfor}).

The five invariant operators are defined by the following expressions
\cite{KRST01}: 
\begin{itemize}
\item The gauge independent operator ${\cal D}_{kin}$ is strictly
one-loop and expresses the renormalization of the local coupling:
\begin{eqnarray}
\label{Dkin}
{\cal D}^ {(1)}_{\rm kin}
 &= & \intd e^ 2\Bigl(
e \frac{\delta} {\delta e} - A^ \mu   \frac{\delta} {\delta A^ \mu}
- \lambda^ \alpha  \frac{\delta} {\delta \lambda^ \alpha} - 
\lambdabar^ \alphadot  \frac{\delta} {\delta \lambdabar^ \alphadot}
\nonumber  \\
 & & \phantom{   \intd e} 
+ 2 \frac {M_\lambda} {e^ 2} \bigl(\dfunc{ } {f} + \dfunc {
}{\fbar}\bigr)
\nonumber \\
 & & \phantom{   \intd e} 
+ Y_ \lambda^ \alpha  \frac{\delta} {\delta Y_\lambda^ \alpha} + 
Y _{\lambdabar}^ \alphadot  \frac{\delta} {\delta Y _{\lambdabar}^ \alphadot}
- c  \frac{\delta} {\delta c} \nonumber 
\\ & & \phantom{   \intd e} 
+ B  \frac{\delta} {\delta B} +  \cbar\frac{\delta} {\delta \cbar} 
 - 2(\xi (x) + \xi)  \frac{\delta}
{\delta \xi} - 2 \chi_\xi  \frac{\delta} {\delta \chi_\xi}\Bigr)\;
\end{eqnarray}
\item The gauge independent operator ${\cal D}^ {\rm sym}_{vV}$ extends
 the  $\brs _\Ga$ invariant operator ${\cal D}_{vV}$ to an 
$\brs_\Ga + r ^{(1)}\delta {\cal S}$-symmetric operator:
\begin{equation}
\label{DvVsym2}
{\cal D}^ {\rm sym}_{Vv} \equiv
{\cal D}^ {(l)}_{Vv} - r^ {(1)}\bigl( 4 {\cal D}^ {(l+1)}_{e} +
8 l ( {\cal N}^ {(l+1)}_V - 8 (l+1)r ^{(1)} \delta {\cal N}^ {(l+2)}_V )
 \bigr) .
\end{equation}
The $\brs_\Ga$ symmetric operator is defined by:
\begin{eqnarray}
  {\cal D}^{(l)}_{Vv} 
& \equiv & \intd
 \biggl( v^ {{(E^{2l})} \mu} \frac {\delta} {\delta V^ \mu} +
\lambda^{{(E^{2l})}\alpha}
\frac {\delta} {\delta \lambdaV^\alpha}
+
\lambdaVbar ^ {{(E^ {2l})}\alphadot}
\frac {\delta} {\delta \lambdaVbar^\alphadot}
 \nonumber \\
& & { } \phantom{\intd}
+ d^ {(E^{2l})}  
\frac {\delta}{\delta \DV}
- i (\epsilon ^ \alpha \chi^ {(E^{2l})}_\alpha - \chibar^ {(E^{2l}) }
_\alphadot \epsilonbar^ \alphadot ) \frac {\delta}{\delta \tilde c} 
\nonumber \\
& & { }\phantom{\intd}
- 2 e^ {2l} (q +m) \frac {\delta }{\delta q} - 2 \bigl(e^{2l}
q^\alpha + 2 \chi^{{(E^{2l})}\alpha}(q+m)\bigr) 
\frac {\delta} {\delta q^ \alpha}
 \nonumber \\
& & { } \phantom{\intd}  -2
\bigl(e^ {2l}(q_F -b ) + 2 f^ {(E^{2l})}(q+m)-  \chi^{{(E^{2l})}\alpha} q_\alpha\bigr)
 \frac {\delta } {\delta q_F}
 \nonumber \\
& & { } \phantom{\intd}
- 2 e^ {2l} (\qbar +m) \frac {\delta }{\delta \qbar} - 2 \bigl(e^{2l}
\qbar^\alphadot +2 \chibar^{{(E^{2l})}\alphadot}(\qbar+m)\bigr) 
\frac {\delta} {\delta \qbar^ \alphadot}
 \nonumber \\
& & { } \phantom{\intd} -2
\bigl(e^{2l}(\qbar_F -b)
 + 2\fbar^ {(E^{2l})}(\qbar+m)-  \chibar^{(E^{2l})}_\alphadot \qbar
 ^\alphadot\bigr) 
 \frac {\delta } {\delta \qbar_F} \biggr) 
\label{DVv}
\end{eqnarray}
The operator
  ${\cal N}^ {(l+1)}_V $ contributes to an anomalous dimension of the
axial vector field and its superpartners:
\begin{eqnarray}
\label{NV}
{\cal N}^ {(l)}_V & = &
 \intd \biggl( e^ {2l}\Bigl( V^ {\mu} \frac {\delta }{\delta V^ \mu} +
 \lambdaV^ {\alpha} \frac {\delta }{\delta \lambdaV^ \alpha} +
 \lambdaVbar^ {\alphadot} \frac {\delta }{\delta \lambdaVbar^ \alphadot} +
 (\tilde D -2 M^2) \frac {\delta }{\delta \tilde D}\Bigr) \nonumber \\
& & \qquad -\frac i 2 V_\mu (\sigma^ \mu \chibar^ {(E ^{2l})})^ \alpha \frac {\delta }{\delta
 \lambdaV^ \alpha} + \frac i2  V_\mu (\chi^  {(E ^{2l})} \sigma^ \mu )^ \alphadot       
\frac {\delta }{\delta \lambdaVbar^ \alphadot}  \nonumber \\
& &  \qquad + 2 \Bigr( V^ \mu v_{\mu}^
 {(E^ {2l})}  + i \lambdaV^ \alpha \chi^ {(E^ {2l})}_\alpha 
- i\chibar^ {(E^ {2l})}_ \alphadot\lambdaVbar^ \alphadot 
 \Bigl)
\frac {\delta }{\delta \tilde D }  \biggr),
\end{eqnarray}
and 
\begin{eqnarray}
\delta {\cal N}^ {(l)}_V
 & = &  \intd e^ {2l} V^ \mu V_\mu \frac {\delta }{\delta \tilde D }\ ;
\end{eqnarray}
 the operator $ {\cal D}_e $ describes a
 redefinition of the coupling $e(x)$ and its superpartners:
\begin{eqnarray}
\label{De}
{\cal D}^ {(l+1)}_e & = & 
 \intd \biggl( e^ {2l+3}\frac {\delta }{\delta e} - 2 \bigl(
\chi^ {(E^ {2l})\alpha} \frac {\delta }{\delta \chi^ \alpha} +
\chibar^ {(E^ {2l})\alphadot} \frac {\delta }{\delta \chibar^
\alphadot}\bigr) \nonumber \\
& & \phantom{ 4 \intd} - 2
\bigl(
(f^ {(E^ {2l})} - M_\lambda l e^ {2l}) \frac {\delta }{\delta f} +
(\fbar^ {(E^ {2l})}- M_\lambda l e^ {2l})
 \frac {\delta }{\delta \fbar}\bigr) \biggr)
\nonumber \\
& & \phantom{ 4\intd}
 - e^{2(l+1)}\Bigl( A^ \mu   \frac{\delta} {\delta A^ \mu}
+ \lambda^ \alpha  \frac{\delta} {\delta \lambda^ \alpha} +
\lambdabar^ \alphadot  \frac{\delta} {\delta \lambdabar^ \alphadot}
- B  \frac{\delta} {\delta B}
\nonumber \\ 
& & \phantom{4\intd - e^ {2(l+1)}}
- Y_\lambda^ \alpha  \frac{\delta} {\delta Y_\lambda^ \alpha} -
Y _\lambdabar^ \alphadot  \frac{\delta} {\delta Y_\lambdabar^ \alphadot}
+ c  \frac{\delta} {\delta c}  -  \cbar\frac{\delta} {\delta \cbar} 
\nonumber \\
& & \phantom{4\intd - e^ {2(l+1)}}
 + 2(\xi (x) + \xi)  \frac{\delta}
{\delta \xi} + 2 \chi_\xi  \frac{\delta} {\delta \chi_\xi}\Bigr)\ .
\end{eqnarray}
In eqs.~(\ref{DVv}),(\ref{NV}) and (\ref{De})
   the components of the  multiplet $E^
{2l}$ are defined by the following expansion:
\begin{eqnarray}
\label{E2ldef}
E^{2 l}(x, \theta, \thetabar )  & = & \bigl({\mbox{\boldmath{$\eta$}}}(x,
\theta, \thetabar ) +{\mbox{\boldmath{$\etabar$}}}(x, \theta,
\thetabar ) + \frac {M_\lambda}{e^2} (\theta^2 + \thetabar^2)\bigr)^ 
{-l} \nonumber  \\
&\equiv &  e^{2l} (x) + \theta^ \alpha \chi^{(E^ {2l})}_\alpha +
 \chibar^{(E^{2 l})}_\alphadot \thetabar^ \alphadot + 
\theta^ 2 f^ {(E^ {2l})} +\thetabar^ 2 \fbar^ {(E^ {2l})}
\nonumber  \\
& & { } 
+ \theta \sigma^ \mu \thetabar v_\mu^ {(E^ {2l})} 
+i \theta^ 2 (\lambdabar^{(E^ {2l})}+
\frac 12 \partial_\mu \chi  \sigma^\mu)
 \thetabar 
\nonumber  \\
& & { }
-i \thetabar^ 2 \theta (\lambda^{(E^ {2l})} +
\frac 12  \sigma^\mu \partial_\mu \chibar^{(E^{2l})} )  
    + \frac 14
\theta^ 2 \thetabar^2  (d^ {(E^ {2l}) }\! - \! \Box e^{2l}) 
\end{eqnarray} 
For constant coupling  one gets:
\begin{eqnarray}
\lim_{E\to e}  f^ {(E^ {2l})} &=  & 
\lim_{E\to e}  \fbar^ {(E^ {2l})} = -l M_\lambda e ^{2l} \ , \nonumber \\
\lim_{E\to e}  d^ {(E^ {2l})} &= & 4 l(l+1) e^ {2l} M_\lambda^ 2 \ ,
\end{eqnarray}
all other $\theta$-components are zero.
\item The three operators corresponding to the field redefinitions are
BRS variations and as such they are gauge dependent:
\begin{eqnarray}
\label{ctvarphi} 
{\cal N}^ {(l)}_\varphi \Ga + \Delta^ {(l)}_{Y,\varphi}
 & \equiv & s_{\Ga} \intd e^{2l} f^ {(l)}_{\varphi}(\tilde \xi)
\bigl(Y_{\varphi_L} \varphi_L  +
Y_{\phibar_L} \phibar_L +  (_{L\to R})  \bigr)  \\
\label{ctpsi} 
{\cal N}^ {(l)}_\psi \Ga + \Delta^ {(l)}_{Y,\psi}
 & \equiv & s_{\Ga}  \intd e^{2l} f^ {(l)}_{\psi}(\tilde \xi)
\bigl(\psi_L Y_{\psi_L}    +
\psibar_L Y_{\psibar_L}  + (_{L\to R})  \bigr)   \\
{\cal N}^ {(l)}_{\psi\varphi} \Ga + \Delta^
{(l)}_{Y,\psi\varphi} & \equiv &
( s_{\Ga} + r^ {(1)} \delta {\cal S})
 \intd \sqrt 2  f^ {(l)}_{\psi\varphi}(\tilde \xi)
\bigl(\chi^ {E^ {(2l)}} \varphi_L Y_{\psi_L}    \nonumber \\
& & \phantom{( s_{\Ga} + r^ {(1)} \delta {\cal S})
 \intd \sqrt 2} + {\chibar^  {E^ {(2l)}}}
\phibar_L Y_{\psibar_L} + (_{L\to R})\bigr)  
\label{ctpsivarphi}
\end{eqnarray}
Explicit expressions are immediately obtained by evaluating the
$\brs _\Ga + r ^{(1)}\delta {\cal S}$-variation. They can be also
found in \cite{KRST01}.
\end{itemize}

\end{appendix}


\end{document}